\documentclass[%
 reprint,
 amsmath,amssymb,
]{revtex4-1}

\usepackage{multirow}
\usepackage{graphicx}
\usepackage{dcolumn}
\usepackage{bm}
\usepackage{breqn}
\usepackage{xcolor}

\begin{document}


\title{Excitation of high-frequency magnon modes in magnetoelastic films by short strain pulses}

\author{Andrei V. Azovtsev}
\author{Nikolay A. Pertsev}
\affiliation{Ioffe Institute 194021 St. Petersburg Russia}

\date{\today}

\begin{abstract}
Development of energy efficient techniques for generation of spin waves (magnons) is important for implementation of low-dissipation spin-wave-based logic circuits and memory elements. A promising approach to achieve this goal is based on the injection of short strain pulses into ferromagnetic films with a strong magnetoelastic coupling between spins and strains.  Here we report micromagnetoelastic simulations of the magnetization and strain dynamics excited in Fe$_{81}$Ga$_{19}$ films by picosecond and nanosecond acoustic pulses created in a GaAs substrate by a transducer subjected to an optical or electrical impulse. The simulations performed via the numerical solution of the coupled Landau-Lifshitz-Gilbert and elastodynamic equations show that the injected strain pulse induces an inhomogeneous magnetization precession in the ferromagnetic film. The precession lasts up to 1 ns and can be treated as a superposition of magnon modes having the form of standing spin waves. For Fe$_{81}$Ga$_{19}$ films with nanoscale thickness, up to seven (six) distinct modes have been revealed under free-surface (pinning) magnetic boundary conditions. Remarkably, magnon modes with frequencies over 1 THz can be excited by acoustic pulses with an appropriate shape and duration in the films subjected to a moderate external magnetic field. This finding shows that short strain pulses represent a promising tool for the generation of THz spin waves necessary for the implementation of high-speed magnonic devices.
\end{abstract}

\maketitle


\setlength{\parindent}{15pt}
\setlength{\parskip}{0pt}

Short strain pulses having duration of a few picoseconds can be created in metals, semiconductors, and insulators by means of a femtosecond optical excitation \cite{Thomsen:1986, Hao:2001D, Hao:2001S, Saito:2003, Muskens:2002, vanCapel:2010, Thevenard:2010}. Such strain pulses give rise to the generation of high-frequency acoustic phonons \cite{Babilotte:2010} and the formation of solitons \cite{Hao:2001S, Muskens:2002, vanCapel:2010}. In magnetic materials with significant magnetoelastic coupling between spins and strains, picosecond strain pulses can induce magnetization precession \cite{Scherbakov:2010, Kim:2015, Deb:2018} and generate spin waves (magnons) with frequencies in the GHz range \cite{Bombeck:2012}. Importantly, spin waves can be employed as information carriers enabling the development of advanced computing technologies with ultralow power consumption in the field of magnon spintronics \cite{Chumak:2015}. Since the wave frequency defines the maximum clock rate of a computing device \cite{Chumak:2015}, it is highly desirable to have a low-dissipation technique for the excitation of magnons with THz frequencies. Furthermore, the magnetization precession was shown to be a source of high-amplitude microwave magnetic field at the nanoscale \cite{Scherbakov:2018}, and even emission of THz electromagnetic waves resulting from the spin dynamics has been demonstrated \cite{Walowski:2016}. Therefore, it is of both fundamental interest and practical significance to achieve the generation of THz spin waves by short strain pulses.
\begin{figure}
    \center{\includegraphics[width=1\linewidth]{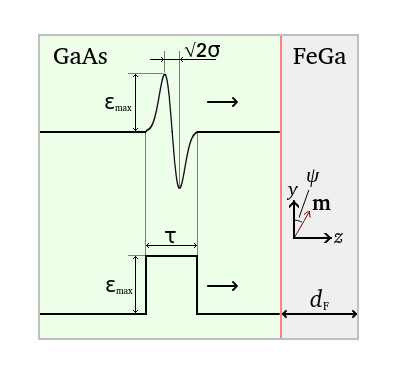}}
    \caption{\label{fig:setup}
  Schematic representation of bipolar and rectangular strain pulses used in simulations of the GaAs/FeGa heterostructure. The time $\tau$ denotes the pulse duration, $\varepsilon_{\text{max}}$ is the maximal strain in the pulse, and $\textbf{m}$ is the unit vector defining the magnetization direction in the ferromagnetic FeGa film with the thickness $d_{\text{F}}$.
  }
\end{figure}

In this paper, we present numerical simulations of the magnetization and strain dynamics induced in ferromagnetic films by acoustic pulses injected from a nonmagnetic substrate. Our \textit{micromagnetoelastic} simulations fully take into account the interplay between spins and elastic strains, which is accomplished via the solution of a system of coupled differential equations for the magnetization and mechanical displacement \cite{APL:2017, PRB:2019}. We consider both picosecond strain pulses created via femtosecond optical excitation and nanosecond pulses generated by a piezoelectric transducer attached to the ferromagnetic heterostructure. The strain-induced magnetization dynamics is studied for films made of FeGa alloy having strong magnetoelastic coupling \cite{Parkes:2013}. Numerical results obtained for FeGa films with nanoscale thickness predict the possibility of non-thermal generation of magnons with frequencies exceeding 1 THz. Similar results are expected for thin films of other ferromagnets exhibiting sufficiently strong magnetoelastic coupling, such as nickel, CoFe alloys, and cobalt ferrite.

The micromagnetoelastic simulations were carried out for a bilayer comprising ferromagnetic (FM) and nonmagnetic (NM) components (Fig. \ref{fig:setup}). The generation of an acoustic pulse was modeled by imparting a time-dependent displacement $u_z(z=0,t)$ to the surface $z=0$ of the NM layer. For picosecond strain pulses corresponding to a femtosecond laser excitation of a metallic film deposited on the rear side of the substrate (upper strain profile in Fig. \ref{fig:setup}) \cite{Scherbakov:2010}, we assumed a Gaussian function $u_z(z=0,t) = u_{\text{max}} \exp{(-t^2/\sigma^2)}$. Such time dependence provides a "bipolar" pulse of the longitudinal strain $\varepsilon_{zz}(z, t) = \partial u_z(z, t)/\partial z$ containing periods of compression and tension characteristic of this technique. The chosen pulse durations of 4 and 40 ps lie on the timescale accessible in the experiments \cite{Hao:2001D, Scherbakov:2010}. To model nanosecond strain pulses created by a piezoelectric transducer attached to the substrate (lower strain profile in Fig. \ref{fig:setup}), a linear variation $u_z(z=0,t) = u_{\text{max}}t/\tau$ during the period $0 \le t \le \tau$ was introduced, which provides the expected rectangular pulse \cite{Peng:2017} with the duration $\tau$.

The formation and propagation of strain pulses in the NM layer was quantified by solving numerically the linear elastodynamic equation for the local displacement $\textbf{u}(z, t)$ \cite{Landau:elasticity}. The terms allowing for non-linearity, dispersion, and scattering with thermal phonons \cite{vanCapel:2010} were neglected because our model calculations are aimed at the description of the strain pulse effect on a thin ferromagnetic film, and because under certain conditions strain pulses retain almost unchanged profiles over very long distances in many crystals \cite{Hao:2001D}. The coupled magnetization and elastic dynamics of the FM layer was described by the Landau-Lifshitz-Gilbert (LLG) equation taking into account the magnetic damping and the elastodynamic equation appended by magnetoelastic terms \cite{APL:2017, PRB:2019}. The LLG equation was written for the unit vector $\textbf{m}(z, t) = \textbf{M}(z, t)/M_s$ defining the local magnetization direction, since the saturation magnetization $M_s$ can be regarded as a constant quantity when the temperature is well below the Curie temperature. The effective field $\textbf{H}_{\text{eff}} = -\partial F/\partial \textbf{M}$ involved in the LLG equation was evaluated with the account of all relevant contributions to the free energy density $F(\textbf{M})$, which result from the exchange interaction, magnetocrystalline anisotropy, magnetoelastic coupling, Zeeman energy, and dipolar interactions between individual spins \cite{PRB:2016}. We considered ferromagnets with cubic paramagnetic phase and described the magnetoelastic contribution $F_{\text{ME}}$ by the relation
\begin{dmath}\label{eq:energy_density}
F_{\text{ME}} = B_1 \left[ (m_x^2-1/3)\varepsilon_{xx} + (m_y^2-1/3)\varepsilon_{yy} \\+ (m_z^2-1/3)\varepsilon_{zz} \right] + B_2 \left[ m_x m_y \varepsilon_{xy} + m_x m_z \varepsilon_{xz} + m_y m_z \varepsilon_{yz} \right],
\end{dmath}
where $B_1$ and $B_2$ are the magnetoelastic coupling constants \cite{Kittel:1949}. Equation (\ref{eq:energy_density}) was also used to determine the magnetoelastic terms in the elastodynamic equation, which are governed by the magnetic contribution $\delta \sigma_{ij} = \partial F_{\text{ME}}/\partial \varepsilon_{ij}$ to mechanical stresses $\sigma_{ij} (i, j = x, y, z)$ in the ferromagnet. These terms enabled us to allow for the backaction of magnetization reorientations on the strain pulse formed in the FM layer, which was ignored in the previous studies of the magnetic dynamics induced by injected acoustic pulses \cite{Linnik:2011, Bombeck:2012, Kim:2012, Kovalenko:2013}. Overall, our rigorous theoretical description of the problem significantly differs from the preceding approximate approaches \cite{Linnik:2011, Bombeck:2012}, which either neglected the exchange interaction and the Gilbert damping \cite{Linnik:2011} or solved the linearized Landau-Lifshitz equation valid only for small-angle magnetization precession \cite{Bombeck:2012} in the absence of damping \cite{Supp3}. It should be noted that short strain pulses represent a mechanical stimulus very different from periodic elastic waves, which were considered in many other studies of elastically excited magnetization dynamics \cite{APL:2017, PRB:2019, Keshtgar:2014, Kamra:2015, Chen:2017}. Indeed, although strain pulse can be represented as a superposition of such waves, the combination of multiple elastic waves with various amplitudes and phases can create quite complex magnetization dynamics very different from those induced by individual monochromatic waves. Furthermore, the specific spatio-temporal profile of the strain pulse inside the FM film differs from that at the NM surface due to its transmission through the NM-FM interface and reflection from the boundary of the FM film. The determination of that profile, which affects the magnetic response of the film, requires numerical simulations, which we performed accounting for all relevant physical effects including the magnetoelastic feedback.

The numerical integration of the LLG equation was carried out with the aid of the projective Runge-Kutta algorithm, while the elastodynamic equation was solved using a finite-difference technique with midpoint derivative approximation. A fixed integration step $\delta t = 0.5$ fs and nanoscale computational cells ($0.1\times0.1\times0.1 ~\text{nm}^3$ \cite{Supp4}) were used in our simulations. We considered both free-surface ($\partial \textbf{m}/\partial z = 0$) and pinning ($m_y = 1$) magnetic boundary conditions for the FM layer \cite{Note1}, which are abbreviated as FBC and PBC below. The free mechanical boundary condition $\sigma_{iz} = 0$ was imposed at the surface of the FM layer, whereas the continuity conditions at the NM$|$FM interface were satisfied automatically due to the introduction of a unified ensemble of computational cells covering the whole NM-FM bilayer. The maximal displacements $u_{\text{max}} \sim 0.0041-7.33$ nm at the surface of the NM layer were chosen to yield a realistic strain amplitude $\varepsilon_{\text{max}} \approx 10^{-3}$ in the acoustic pulse \cite{Peng:2017, Peronne:2017}. Since galfenol films are successfully grown on GaAs substrates \cite{Parkes:2013}, the NM layer was assumed to be made of GaAs. Accordingly, we attributed the elastic stiffnesses $c_{11} = 1.1877\times10^{12}$ dyne cm$^{-2}$, $c_{12} = 0.5372\times10^{12}$ dyne cm$^{-2}$, $c_{44} = 0.5944\times10^{12}$ dyne cm$^{-2}$ and the mass density $\rho_{\text{NM}} = 5.3169$ g cm$^{-3}$ to the NM layer \cite{CRC}. For the FM layer, the following material parameters characterizing Fe$_{81}$Ga$_{19}$ alloy were used in the simulations: exchange stiffness $A_{\text{ex}} = 2.1\times10^{-6}$ erg cm$^{-1}$ \cite{Scherbakov:2019}, cubic anisotropy constant $K_1 = 2.65\times10^5$ erg cm$^{-3}$ \cite{Scherbakov:2019}, magnetoelastic coefficients $B_1 = -0.9\times10^8$ erg cm$^{-3}$ and $B_2 = -0.8\times10^8$ erg cm$^{-3}$ \cite{Restorff:2012}, saturation magnetization $M_s = 1472$ emu cm$^{-3}$ \cite{Scherbakov:2019}, Gilbert damping parameter $\alpha = 0.017$ \cite{Jager:2013}, elastic moduli $c_{11} = 1.62\times10^{12}$ dyne cm$^{-2}$, $c_{12} = 1.24\times10^{12}$ dyne cm$^{-2}$, $c_{44} = 1.26\times10^{12}$ dyne cm$^{-2}$ and mass density $\rho_{\text{F}} = 7.8$ g cm$^{-3}$ \cite{Basantkumar:2006}. External magnetic field $H_z = 5$ kOe was applied in the perpendicular-to-plane direction to facilitate the excitation of the magnetization precession by the longitudinal strain $\varepsilon_{zz}$. An in-plane field $H_y = 2$ kOe was also introduced to stabilize the single-domain state. In the presence of such magnetic fields, the initial magnetization direction has an elevation angle of $\psi \approx 14^{\circ}$ in the galfenol film with FBC, which is homogeneously magnetized initially (see Fig. \ref{fig:setup}). Under PBC, the equilibrium magnetization $\textbf{M}(z, t = 0)$ assumes a non-uniform distribution across the film, which was determined by preliminary runs.
\begin{figure}
  \begin{minipage}{0.9\linewidth}
    \center{\includegraphics[width=1\linewidth]{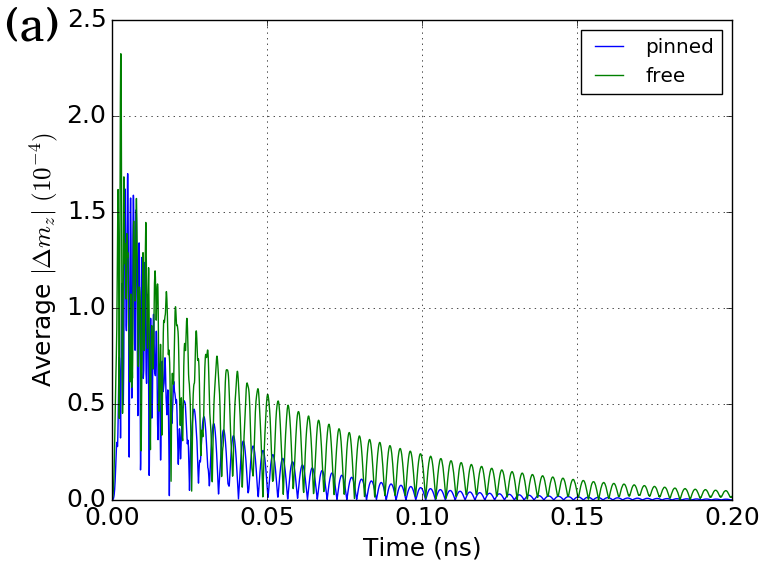}}
  \end{minipage}
  \begin{minipage}{0.9\linewidth}
    \center{\includegraphics[width=1\linewidth]{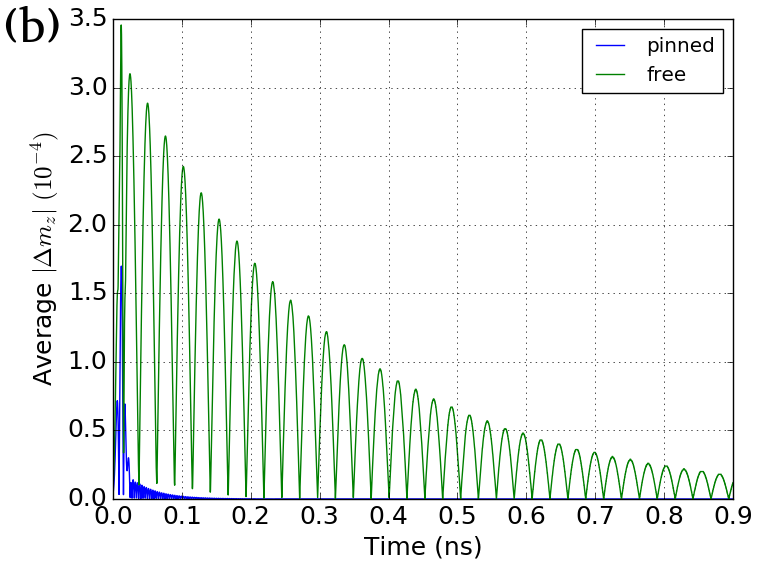}}
  \end{minipage}
  \begin{minipage}{0.9\linewidth}
    \center{\includegraphics[width=1\linewidth]{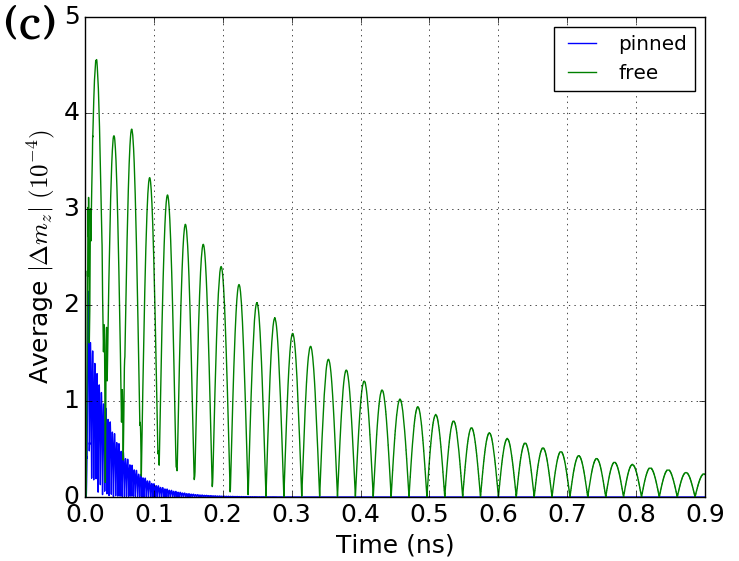}}
  \end{minipage}
    \caption{\label{fig:precession}
  Magnetization precession excited in the 8-nm-thick Fe$_{81}$Ga$_{19}$ film by bipolar strain pulses having duration of 4 ps (a) and 40 ps (b) and by 1-ns-long rectangular pulse (c). Presented graphs show time dependences of the absolute value of the magnetization direction cosine $|m_z|$ averaged over the film thickness $d_{\text{F}}$. Green and blue curves correspond to free-surface and pinning boundary conditions, respectively. The maximal strain $\varepsilon_{\text{max}}$ in all pulses equals $1.55\times10^{-3}$. Dynamical temporal evolutions of spatial profiles of longitudinal strain $\varepsilon_{zz}$ and direction cosines $m_i$ in 8-nm-thick film are shown in multimedia views: https://doi.org}
\end{figure}

The simulations confirmed that the aforementioned displacements $u_z(z = 0, t)$ imparted to the GaAs surface create strain pulses with the shapes depicted in Fig. \ref{fig:setup}. After crossing the GaAs layer, the strain pulse penetrates into the galfenol film (transmittance with respect to energy is about 97\% due to similar acoustic impedances of GaAs and FeGa), where it induces a non-uniform magnetization precession. The backaction of such precession on strain state of the FM film manifests itself in the generation of weak transverse elastic waves, which involve shear strains $\varepsilon_{xz}$ and $\varepsilon_{yz}$ not present in the injected acoustic pulse. It should be noted that the 1-ns-long rectangular pulse formed in the NM substrate actually transforms into a much shorter strain pulse acting on the FM film. Such transformation is caused by the reflection of the pulse from the free boundary of the film, which occurs with the phase shift of 180$^{\circ}$, and its consequent destructive interference with the incoming pulse. This yields a rectangular pulse with the duration $\tau_{\text{F}} = 2d_{\text{F}}/c_{\text{L}}$, where $c_{\text{L}} = \sqrt{c_{11}/\rho_{\text{F}}}$ is the velocity of the longitudinal elastic waves in FM layer. Thus, any incoming rectangular strain pulse longer than $\tau_{\text{F}}$ (3.5 ps in 8-nm-thick FeGa film and 7 ps in 16-nm-thick one) will create a shorter pulse with the duration $\tau = \tau_{\text{F}}$ depending on the thickness of the FM film. Remarkably, this feature renders possible to generate picosecond rectangular strain pulses with a modified spectrum in FM nanolayers even by long electrical pulses applied to a piezoelectric transducer attached to a NM substrate.

To characterize the magnetization precession in FeGa, we determined the deviation $\Delta m_z$ of the direction cosine $m_z$ from its equilibrium value and calculated the average deviation $\langle |\Delta m_z| \rangle$ in the film. Figure \ref{fig:precession} shows time dependences of $\langle |\Delta m_z| \rangle$ in the 8-nm-thick Fe$_{81}$Ga$_{19}$ film subjected to bipolar and rectangular strain pulses with different durations but the same magnitude $\varepsilon_{\text{max}} = 1.55\times10^{-3}$. It can be seen that the precession amplitude decreases with time due to considerable magnetic damping inherent to the galfenol film. Depending on the magnetic boundary conditions, pulse shape and duration, magnetization precession lasts from $\sim 0.1$ ns to $\sim 1$ ns, because those factors determine the exact spectral composition of spin signal, which contains specific magnon modes with different lifetimes. The inspection of Fig. \ref{fig:precession} demonstrates that the precession lifetime increases markedly in the film with FBC. This feature is due to the presence of spatially uniform mode of the magnetization precession (see below), which has lower frequency and, therefore, decays slower than other modes. Furthermore, the comparison of panels (a), (b), and (c) in Fig. \ref{fig:precession} reveals that longer strain pulses excite more long-living precession in the film with FBC, which can be explained by small amplitude of the uniform mode excited by the 4-ps-long pulse. The shape of the strain pulse mostly affects the amplitude of the excited precession under FBC. Namely, rectangular pulse of duration $\tau = 1$ ns excites magnetization precession with larger amplitude than both considered bipolar pulses. It should be noted that the precession amplitude is also sensitive to magnetic boundary conditions, generally being significantly smaller under PBC due to the absence of the uniform mode (compare blue and green curves in Fig. \ref{fig:precession}). Interestingly, ultrashort bipolar pulse with duration $\tau \approx 4$ ps initially creates magnetization precession of similar magnitude under both boundary conditions. This situation, which is different from the one appearing under two longer pulses, is due to the prevalence of a non-uniform high-frequency mode in the magnon spectrum generated by such an ultrashort pulse, which has a maximum of excitation at a high frequency of about 0.5 THz (see Fig. 3 and Table I below).

\begin{figure}[t]
  \begin{minipage}{0.79\linewidth}
    \center{\includegraphics[width=1\linewidth]{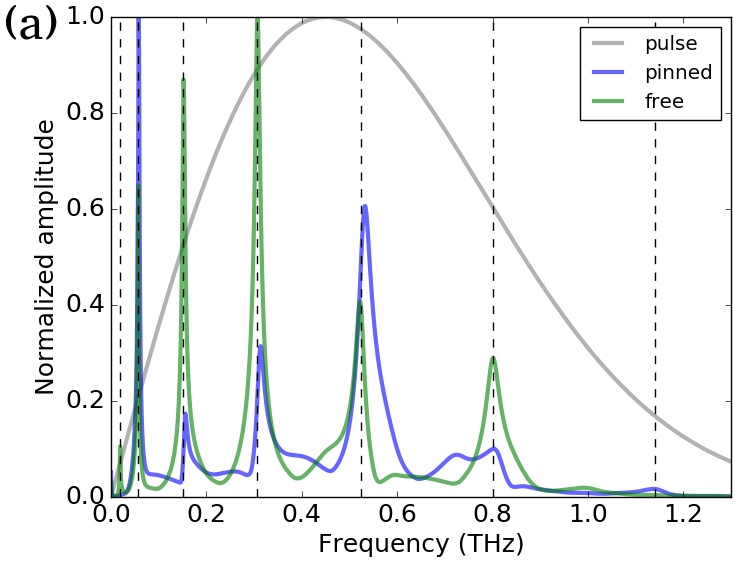}}
  \end{minipage}
  \begin{minipage}{0.79\linewidth}
    \center{\includegraphics[width=1\linewidth]{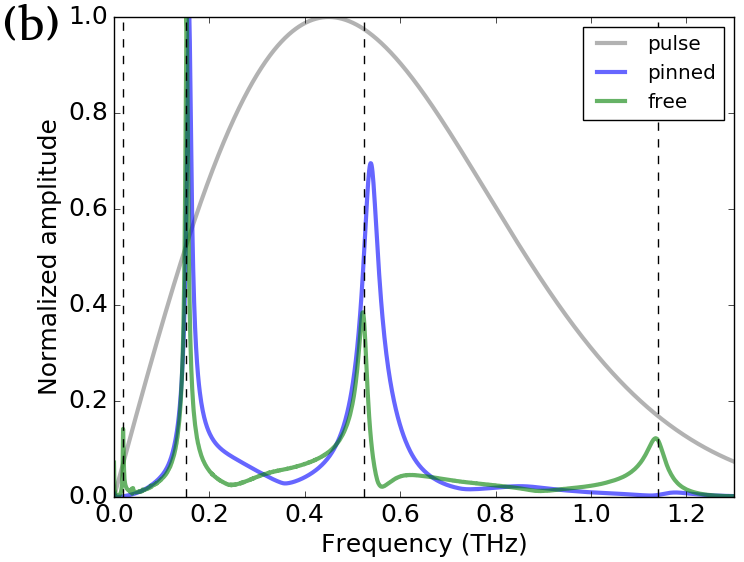}}
  \end{minipage}
  \begin{minipage}{0.79\linewidth}
    \center{\includegraphics[width=1\linewidth]{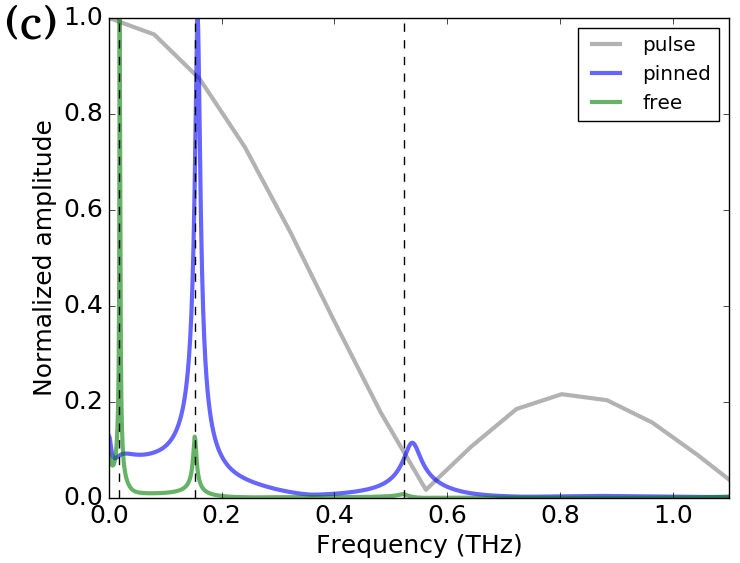}}
  \end{minipage}
    \caption{\label{fig:spectrum} Frequency spectra of standing spin waves excited in Fe$_{81}$Ga$_{19}$ films by 4-ps-long bipolar strain pulse [(a), (b)] and by 1-ns-long rectangular one (c). The spectra show the thickness-averaged amplitudes $\langle A\rangle(\nu)$ of the Fourier transforms of the time dependences $\Delta m_z(t)$, which are normalized by the height of the strongest peak revealed at the given pulse duration, film thickness, and magnetic boundary conditions. Green and blue curves correspond to FBC and PBC, respectively. Vertical dashed lines indicate the frequencies $\nu_n(k_n)$ calculated with the aid of Eq. (\ref{eq:dispersion}). Grey curve (solid) shows the spectrum of the strain pulse. The film thickness equals 16 nm (a) and 8 nm [(b), (c)].}
\end{figure}

Spatial distribution of the strain-induced magnetization precession across the film thickness is strongly inhomogeneous and can be treated as a superposition of standing spin waves with different wavelengths. To determine the frequency spectrum of magnetic excitations, we calculated the Fourier transforms of the time dependences $\Delta m_z(t)$ registered in each computational cell inside the galfenol film. By averaging the frequency-dependent amplitudes $A(\nu, z)$ of such transforms over all cells, we obtained the spectra $\langle A \rangle(\nu)$ for the Fe$_{81}$Ga$_{19}$ films of different thickness subjected to bipolar and rectangular strain pulses (see Fig. \ref{fig:spectrum}). It was found that such spectra may contain up to seven (six) well-defined peaks under FBC (PBC). To determine the origin of these peaks, we employed the dispersion relation of exchange-dominated spin waves with the wavevector $\textbf{k}$ perpendicular to the film plane \cite{Shihab:2015}
\begin{dmath}\label{eq:dispersion}
\nu(k) = \frac{1}{2\pi}\sqrt{F_{\theta\theta}F_{\phi\phi}/M_s^2 + (F_{\theta\theta}+F_{\phi\phi})Dk^2/M_s + D^2k^4},
\end{dmath}
where $F_{\theta\theta}$ and $F_{\phi\phi}$ are the second derivatives of the free energy density $F(\textbf{M})$ with respect to the polar and azimuthal angles $\theta = 90^{\circ}-\psi$ and $\phi$ defining the magnetization direction in the film (see Fig. \ref{fig:setup}), $D = 2\gamma A_{\text{ex}}/M_s$ is the spin wave stiffness constant, and $\gamma$ is the electron's gyromagnetic ratio. Using Eq. (\ref{eq:dispersion}) with the involved energy density $F(\textbf{M})$ calculated with the account of contributions associated with the Zeeman energy, cubic magnetocrystalline anisotropy, and demagnetization field, we determined the frequencies $\nu_n$ of standing spin waves with the wavenumbers $k_n = n\pi/d_{\text{F}}~(n = 0, 1, 2,...$) depending on film thickness $d_{\text{F}}$.
\begin{table*}[t]
    \centering
    \begin{tabular}{|c|c|c||c|c|c|c|c|c|c|}
         \hline
         \multirow{2}*{$d_{\text{F}}$} & \multirow{2}*{$\tau$} & \multirow{2}*{BC} & \multicolumn{7}{c|}{Mode number $n$, frequency in GHz, and relative amplitude} \\
         \cline{4-10}
          & & & 0 & 1 & 2 & 3 & 4 & 5 & 6\\
         \hline
         \multirow{6}*{8 nm} & \multirow{2}*{4 ps} & PBC & - & 157 (0.01417) & 538 (0.00983) & 1177 (0.00013) & & &\\
         \cline{3-10}
          & & FBC & 19 (0.0043) & 152 (0.03047) & 522 (0.0117) & 1136 (0.00373) & & &\\
         \cline{2-10}
          & \multirow{2}*{40 ps} & PBC & - & 156 (0.00369) & & & & &\\
         \cline{3-10}
          & & FBC & 19 (0.36851) & 152 (0.01102) & & & & &\\
         \cline{2-10}
          & \multirow{2}*{1 ns} & PBC & - & 157 (0.02979) & 538 (0.00343) & 1182 (0.00014) & & &\\
         \cline{3-10}
          & & FBC & 19 (0.52766) & 152 (0.06766) & 522 (0.00415) & 1137 (0.00340) & & &\\
         \hline
         \hline
         \multirow{6}{*}{16 nm} & \multirow{2}*{4 ps} & PBC & - & 58 (0.03187) & 156 (0.00553) & 314 (0.01) & 533 (0.01932) & 803 (0.00320) & 1137 (0.00054) \\
         \cline{3-10}
          & & FBC & 19 (0.00528) & 57 (0.03268) & 152 (0.04383) & 307 (0.05021) & 522 (0.02047) & 801 (0.01455) & 1122 (0.00022)\\
         \cline{2-10}
          & \multirow{2}*{40 ps} & PBC & - & 58 (0.999) & 155 (0.00147) & 313 (0.00021) & 530 (0.00015) & &\\
         \cline{3-10}
          & & FBC & 19 (0.68511) & 57 (1.0) & 152 (0.01545) & 307 (0.00056) & & & \\
         \cline{2-10}
          & \multirow{2}*{1 ns} & PBC & - & 58 (0.33064) & 156 (0.01226) & 314 (0.00587) & 533 (0.00668) & 803 (0.00124) & 1141 (0.00048) \\
         \cline{3-10}
          & & FBC & 19 (0.90638) & 57 (0.34340) & 152 (0.10340) & 307 (0.03277) & 522 (0.00723) & 802 (0.00536) & 1132 (0.00019)\\
         \hline
    \end{tabular}
    \caption{Characteristics of magnon modes excited in Fe$_{81}$Ga$_{19}$ films by short acoustic pulses. Amplitudes $\langle A\rangle(\nu_n)$ of all modes are normalized by the maximum amplitude $\langle A\rangle_{\text{max}}$ revealed in our simulations, which corresponds to the magnon mode with $n = 1$ excited by the 40-ps-long bipolar pulse in the 16-nm-thick film. Data in the cells is given in the format "frequency of the peak in GHz (relative amplitude)".}
    \label{tab:spectra}
\end{table*}
The calculated frequencies $\nu_n(k_n)$ are shown by vertical dashed lines in {Fig. \ref{fig:spectrum}}. It can be seen that the peaks of the average amplitude $\langle A \rangle(\nu)$ are situated at frequencies agreeing well with $\nu_n(k_n)$. This result proves that such peaks are caused by the excitation of magnon modes having the form of standing spin waves. The frequencies and relative magnitudes $\langle A\rangle(\nu_n)/\langle A \rangle_{\text{max}}$ of the revealed magnons are summarized in Table \ref{tab:spectra}, which demonstrates that even the modes with $n = 6$ can be excited by acoustic pulses. The fundamental mode with $n = 0$, which corresponds to the coherent magnetization precession in the film ($k_0 = 0$), appears under FBC only, because PBC do not allow the excitation of such precession. Therefore, the mode with $n = 1$ has the lowest frequency in the magnon spectra of ferromagnetic films with PBC. As might be expected, the magnetic boundary conditions only weakly influence the frequencies of magnon modes with $n \ge 1$, which are governed by the film thickness and the mode number $n$. However, in general they strongly affect the amplitudes of such modes due to different restrictions imposed on the magnetization at the film surfaces (see Table \ref{tab:spectra}). Magnon modes with the highest frequencies exceeding 1 THz were revealed in the films subjected to the 4-ps-long bipolar acoustic pulse and the 1-ns-long rectangular one (see Table \ref{tab:spectra}). Bipolar strain pulse with duration $\tau \approx 40$ ps is inefficient for the excitation of THz modes, because the Fourier spectrum of such pulse does not contain significant components at THz frequencies.

The inspection of Table \ref{tab:spectra} shows that, among the THz-frequency magnon modes, the largest relative amplitude $\langle A \rangle(\nu_n) /\langle A \rangle_{\text{max}} \cong 0.0037$ has the mode with $n = 3$ excited by the 4-ps-long bipolar pulse in the 8-nm-thick Fe$_{81}$Ga$_{19}$ film with FBC [Fig. \ref{fig:spectrum}(b)]. The second largest amplitude $\langle A\rangle(\nu_n) /\langle A \rangle_{\text{max}} \cong 0.0034$ corresponds to the same mode generated by the 1-ns-long rectangular pulse in the same film with FBC. Under PBC, the amplitudes of THz modes drop drastically in the 8-nm-thick film and become similar to those excited in the 16-nm-thick one. Thus, the generation efficiency of THz magnons strongly depends on film thickness, magnetic boundary conditions, and pulse duration \cite{Supp2}.

In summary, we carried out the numerical modeling of the magnetoelastic dynamics induced in FM films by short strain pulses generated in the underlying NM substrate with the aid of an optical or electrical stimulus. Our rigorous treatment significantly improves the theoretical description of the magnetic dynamics induced by acoustic pulses in comparison with the preceding approximate approaches \cite{Linnik:2011, Bombeck:2012}. The simulations demonstrated the excitation of inhomogeneous magnetization precession with the duration ranging from $\sim0.1$ ns to $\sim1$ ns in galfenol films. The Fourier analysis of the simulation results revealed that such precession can be treated as a superposition of magnon modes having the form of standing spin waves. Similar modes are expected to form in thin films of other ferromagnets exhibiting sufficiently strong magnetoelastic coupling, such as nickel, CoFe alloys, and cobalt ferrite. Remarkably, even the modes with frequencies over 1 THz can be efficiently excited in thin galfenol films subjected to a magnetic field of only a few kOe. This important result demonstrates that an energy-efficient generator of THz spin waves, which is highly desirable for the development of advanced computing technologies in the field of magnon spintronics, can employ short strain pulses created by optical or electrical impulses.

We are grateful to A. V. Scherbakov and T. L. Linnik for critical reading of the manuscript and their valuable comments. The work was supported by the Foundation for the Advancement of Theoretical Physics and Mathematics "BASIS".

\bibliography{References}

\begin{thebibliography}{39}
\bibitem [{Sup({\natexlab{b}})}]{Supp4}%
  \BibitemOpen
  \href@noop {} {} ({\natexlab{b}}),\ \bibinfo {note} {{Such} small size of
  computational cells was chosen solely to achieve high accuracy of numerical
  calculations aimed at obtaining a smooth solution.}\BibitemShut {Stop}%
\bibitem [{Sup({\natexlab{c}})}]{Supp2}%
  \BibitemOpen
  \href@noop {} {} ({\natexlab{c}}),\ \bibinfo {note} {{The} generation of
  magnons may also be affected by the roughness of the FM$|$NM interface, which
  modifies the magnetic boundary condition and the pulse transmission across
  the interface. However, the presence of such roughness should not change our
  predictions qualitatively, because we found that short strain pulses can
  generate THz magnon modes under both FBC and PBC.}\BibitemShut {Stop}%
\end{thebibliography}
\end{document}